\documentstyle[prl,aps]{revtex}
\begin{document}
\wideabs{
\title{Chiral Multiplets Versus Parity Doublets 
in Highly Excited Baryons}

\author{Thomas D.~Cohen$^a$ and Leonid Ya. Glozman$^b$}

\address{a.  Department of Physics, University of Maryland, 
College Park,
20742-4111\\
b. Institue for Theoretical Physics, University of Graz, 
Universit\"atsplatz 5, A-8010 Graz, Austria}

\maketitle

\begin{abstract}

It has recently been suggested that the  parity
doublet structure seen in the spectrum of highly excited
baryons may be due to effective chiral restoration for
these states. We 
argue how the idea of chiral symmetry restoration  high in
the spectrum is consistent with the concept of quark-hadron duality.
If chiral symmetry is effectively
restored for highly-lying states, then the baryons should
fall into representations of $SU(2)_L \times SU(2)_R$ that
are compatible with the given parity of the states - the parity-chiral
multiplets. We classify
all possible parity-chiral multiplets: 
(i) $(1/2,0) \oplus (0, 1/2)$ that contain parity doublet for 
nucleon spectrum,
(ii) $(3/2,0) \oplus (0, 3/2)$ consists of the
parity doublet for delta spectrum,
(iii) $(1/2,1) \oplus (1, 1/2)$  contains one
parity doublet in the nucleon spectrum and one parity doublet
in the delta spectrum of the same spin that are degenerate in
mass.  
Here we show that the available spectroscopic data for 
nonstrange baryons in the $\sim$ 2 GeV range is consistent
with all possibilities, but the approximate degeneracy
of parity doublets in nucleon and delta spectra support
the latter possibility
 with excited baryons approximately falling into 
$(1/2,1) \oplus (1, 1/2)$ representation of $SU(2)_L \times SU(2)_R$ 
with approximate degeneracy between positive and negative 
parity $N$ and $\Delta$ resonances of the same spin.


\end{abstract} }

\medskip

\section{Introduction}

It is believed that ultimately QCD, the underlying theory
 of the strong interaction, can explain all of the features
 of hadronic physics.  However, the subject remains of 
considerable importance since predictions of hadronic phenomena 
from QCD remains computationally intractable for many problems
 despite advances made in lattice gauge theory.  Moreover,
 the manner that the underlying QCD degrees of freedom transmute
 into the observable hadronic degrees of freedom is subtle and
 complex and of great intellectual interest.  One interesting
 feature of hadronic physics is the appearance of approximate parity 
doublets for highly excited baryons (baryons with a mass 
of $\sim$ 2 GeV and above).  Recently it has been suggested 
that these parity doublets can be explained by an effective
 restoration of chiral symmetry for these highly excited 
states\cite{G1}.

One feature of QCD that is well understood is that the theory
 possess an approximate  $SU(2)_L \times SU(2)_R$ symmetry
 (which becomes exact as the current quark mass goes to zero)
 and that this symmetry is spontaneously broken.  Clearly, in
 the absence of both explicit and spontaneous symmetry 
breaking all hadronic states would fall into chiral
 multiplets and each multiplet would have both positive 
and negative parity states.  For example, in the meson
 sector the mass difference between the positive parity $\sigma$ 
 and the negative parity $\pi$ is entirely due 
to chiral symmetry breaking.  One possible way to understand 
the near degeneracy between highly excited baryons of different
 parities is to suggest that for states in this regime there
 is  some type of  effective chiral symmetry 
restoration.  Although this has been refered to as a phase 
transition \cite{G1} it can be rephrased in the following way.
   The essential conjecture is simply that as one goes up 
in excitation energy in the baryon spectrum the role of chiral 
symmetry breaking  in determining spectral properties
 diminishes to the point where the states act to good 
approximation as though there were no symmetry breaking 
effects.

A natural language to consider this phenomenon is via correlation 
functions of currents (interpolating fields) constructed from 
quark and gluon operators and carrying the quantum numbers of 
baryons as is done in both QCD sum rule \cite{SR} and in lattice
 QCD \cite{LATTICE}. The correlation functions at high
space-like momenta are naturally expressed in terms of an
operator product expansion (OPE) \cite{OPE} which is the
basis of QCD sum-rule calculations.
Thus, the correlation function
at high space-like momenta is dominated by the perturbative
contributions; towards smaller momenta it is dominated
by condensates (vacuum expectation values of composite operators), 
whose contributions are governed by 
inverse powers of momenta  relative to
 the contribution of the three free quark propagators. 
The correlation function in the deep Euclidean region
can be linked to the
imaginary parts of the correlators in the time-like region
via dispersion relations; one integrates over a spectral
 function that is the square of the amplitude that the current
 creates a state of given mass squared.
 The currents can be constructed to have well defined transformation 
properties under $SU(2)_L \times SU(2)_R$ transformations. 
 Because of asymptotic freedom it is clear that the correlation
 functions at asymptotically high momentum can be calculated 
directly in perturbation theory.  However in perturbation 
theory there is no spontaneous symmetry breaking. The spectrum
at asymptotically high mass is connected to only the
asymptotically high momentum in correlator.
Thus one
 sees immediately that 
the chiral $SU(2)_L \times SU(2)_R$ symmetry must be manifest
(explicit) 
 in the spectrum at asymptotically high 
masses (i.e. this spectrum must be  insensitive to the effects of chiral
 symmetry breaking).  
The spectral density at asymptotically high mass associated with
 a particular current is identical to the spectral density for 
a chirally rotated current, {\it i.e.} the states form chiral
 mutliplets.  Although the states form chiral multiplets, 
asymptotically high in the spectrum they cannot be identified 
with hadrons, because  as one increases the mass, the spectrum of 
resonances becomes increasingly dense and the widths 
should not decrease.  Ultimately the resonances  overlap 
to the point at which it is no longer meaningful to 
identify part of the spectral density with a given hadronic
 resonance.  Indeed, it is this structure of dense overlapping
 resonances that allows the hadronic spectrum to approach the
 perturbative QCD continuum that naively is associated with 
multi-quark states.

In terms of this language, the conjecture that there is 
effective chiral restoration in the spectrum of highly 
excited baryons can be understood as follows: as one goes
 up in excitation energy the effects of chiral symmetry
 breaking on the spectrum must diminish as one approaches 
the perturbative regime.  
 The conjecture, then, is simply that the effect of chiral 
symmetry breaking cuts off low enough in the spectrum that
 isolated hadronic resonances are still distinct.  This in
 turn means that chiral symmetry breaking effects become 
negligible in these correlation functions  than at least some other 
nonperturbative (but chiral invariant)  and perturbative effects 
that are responsible for the overal baryon mass high in the spectrum
 (in particular the effects responsible for the formation of hadronic
 resonances which intuitively are related to confinement).

In the next section we show how the concept of quark hadron
duality allows one to expect the effective chiral symmetry
restoration high in the spectrum. The third section is devoted
to a classification of the possible parity-chiral multiplets.
In the fourth section we discuss an alternative possibility
for parity doubling, namely $U(1)_V \times U(1)_A$ restoration,
and show that $U(1)_V \times U(1)_A$ restoration cannot explain
parity doubling unless simultaneously the chiral symmetry 
$SU(2)_L \times SU(2)_R$ is also restored. 
In the fifth section we will review the data on highly 
excited baryonic resonances and show that the pattern of 
excitations is such that the states can be interpreted as 
falling into parity-chiral multiplets. In the final section we
compare with other approach and conclude the present study.

\section{Why should one expect chiral symmetry restoration
high in the spectrum}

In this section we argue how the concept of quark-hadron
duality suggests chiral symmetry restoration
high in the spectrum.
The phenomenon of quark-hadron duality \cite{WEINBERG} is
well established in many processes, e.g. in $e^+e^- \rightarrow
hadrons$, where we have a direct experimental access to
creation of the quark-antiquark pair by the electromagnetic
current. According to this concept, the spectral density
$\rho(s)$ (perhaps appropriately smeared) at the very large $s$ should 
be dual to the polarization operator calculated at the free
quark loop level (up to perturbative corrections). For the
process $e^+e^- \rightarrow hadrons$ the "asymptotic regime"
sets in approximately at $s \sim 2-3$ GeV$^2$ (within the light
flavours sector). The physical picture that is behind such a duality is
rather simple. At large $s$ the convertion of the virtual photon
into quark-antiquark pair is happened at the very short distances
 between  the quark and antiquark 
  (or during very small time interval)         and so this
stage is described by perturbative QCD. Materialization of these
quark and antiquark into physical hadrons happens at the second
stage of the process, where the quark and antiquark are quite far
from each other and the nonperturbative QCD phenomena are important here.
 These nonperturbative phenomena , however,cannot significantly
affect the full (inclusive) transition rate (spectral density)
which is determined at the first stage of the process.

In the case of baryons, unfortunately, there are no experimentally
accessible currents that can create three quarks at some space-time
point and connect them to baryons. Nevertheless, one can construct
such currents theoretically and these currents are widely used in
QCD sum rules or lattice calculations to extract properties of
low-lying baryons directly from QCD. The quark-hadron duality
applied to the present case, would mean that in the asymptotically
high part of the baryon spectrum the baryon spectral density
should be dual to the one which is calculable in perturbation theory;
hence the chiral symmetry should be manifest in the spectral
density, because there is no chiral symmetry breaking in
perturbation theory.

Consider, as an example, the two-point correlator
of the Ioffe current \cite{IOFFE}
$\eta = \epsilon_{abc}\left ( {u^a}^TC\gamma_\mu u^b \right )
\gamma_\mu \gamma_5 d^c$, {\it i.e.} one of the currents that
couples to isodoublet $J=1/2^+$ and $J=1/2^-$ baryons. This
correlator contains chiral even and odd terms

\begin{eqnarray}
\Pi (q) & = & i \int d^4x e^{iqx} \langle  0 
|T\left ( \eta(x), \bar \eta(0) \right)
| 0 \rangle  \nonumber \\
& = &\Pi^{even}(q^2)q_\mu \gamma^\mu + \Pi^{odd}(q^2),
\label{corr}
\end{eqnarray}

\noindent
that behave
differently under discrete chiral transformations of the 
form $\exp (i \pi\gamma_5{\vec \tau} \cdot \hat{n}/2)$ 
(with arbitrary $\hat{n}$); while the former is invariant under this 
transformation , the latter one switches sign.

In the deep space-like domain $q^2 < 0$, where the language of
quarks and gluons is adequate, the OPE up to dimension 
dim=4 operators is \cite{IOFFE,RUB}

\begin{eqnarray}
\Pi^{odd}(q^2) & = &
 -\frac{1}{4\pi^2} \langle \bar q q \rangle q^2 \ln (-q^2) \,+ \, \cdots,
 \label{OPEO} \\
 \Pi^{even}(q^2) & = & 
 \frac{\ln(-q^2)}{32\pi^2}\left(\frac{q^4}{2\pi^2}  
 + \langle\frac{\alpha_s}{\pi} G^a_{\mu \nu}
G^a_{\mu \nu} \rangle \right) \, + \, \cdots .
 \label{OPEE}
 \end{eqnarray}

\noindent
In these equations the first term in the chiral even part
of the correlator represents the zeroth order perturbative
contribution, {\it i.e.} propagation of  three free quarks
from the point $0$, where they are created by the current, to
the point $x$, where they are annihilated by the same current.
The second term of (\ref{OPEE}), which is the contribution of
the gluon condensate, parametrizes soft nonperturbative gluonic
effects.  
Unlike  $\langle \bar q q \rangle$ , the gluon operator
$\langle\frac{\alpha_s}{\pi} G^a_{\mu \nu}
G^a_{\mu \nu} \rangle$ is not a chiral order parameter \cite{SHIFMAN}. 
Higher order
 perturbative corections (containing logarithmic contributions) 
and the contribution of the higher order chirally even
condensates (which are suppressed by powers of $1/q^2$) and that do not 
break the discrete chiral symmetry are not included in eq.~(\ref{OPEE}). 
In contrast, the chiral odd contribution given in eq.~(\ref{OPEO}) 
has no purely perturbative part.  It consists exclusively of
 contributions proportional to the various chirally odd condensates. 
 In the expression (\ref{OPEO}) only the lowest dimension
(dim=3) quark condensate is shown explicitly. The contribution of
other chirally odd condensates of higher dimension
are suppressed by powers of $1/q^2$.

The spectral density, $\rho(s)$, is
proportional to the imaginary part of the correlator in the time-like
region, $ s=q^2 > 0$.
The spectral density parameterizes the amplitude for the current to
 create a baryon state with mass of $s^{1/2}$ and hence provides direct 
 information about the spectrum. While it is not trivial to calculate
  $\rho(s)$ directly, it is straightforward to analytically continue 
  the truncated OPE expansion
from the deep Euclidean domain to the time-like region. While such a 
procedure introduces ambiguities, the ambiguities are suppressed as 
one goes asymptotically high in the spectrum \cite{SH}.  In the present 
context, the significant point is that eqs.~(\ref{OPEE}) and (\ref{OPEO})
 imply that at sufficiently large space-like 
 $q^2$, $C \Pi^{even}(q^2) \gg \Pi^{odd}(q^2)$,
where the dimensional constant C (with dimension MeV) is needed to make
a comparison meaningful.
  Analytically continuing this to the large time-like region implies 
  that for sufficiently large $s$, $C \rho^{even}(s) \gg \rho^{odd}(s)$.  
This in turn implies the spectrum is chirally even, {\it i.e. invariant 
under the discrete chiral transformation}.  However, a stronger 
constraint can be found for the high lying spectrum. Up to dimension 4,
 the chirally-even correlator is invariant under more than the discrete 
 chiral rotations used to define ``even'' and ``odd'' but under arbitrary
  chiral rotations.  This can be seen explicitly from eqn.~(\ref{OPEE})
   which at this order is independent of all chirally active condensates. 
    Thus analytically continuing to large time-like $q^2=s$ one concludes 
    the asympotic spectral function is not only chirally even under the
     discrete transformation but is chirally invariant under arbitrary
      chiral transformations. Summarizing, {\it even if the chiral symmetry
is strongly broken in the vacuum (and hence in the low-lying states),
one should expect that the effects of chiral symmetry breaking become
unimportant for high-lying spectrum.} This is a simple consequence
of the concept of quark-hadron duality.

The Ioffe current that was  considered in the example above
belongs to the  $(1/2,0) \oplus (0,1/2)$ representation of
the chiral group. However, one can also construct the currents
that transform according to the other representations of
this group \cite{CJ}.

\section{Classification of the parity-chiral multiplets}

  Regardless of how plausible one views the
 {\it a priori} arguments above, it is useful to
 see whether the conjecture is consistent at the phenomenological
 level with the spectroscopy of highly excited baryons. 
 In particular, although the principal motivation behind 
this conjecture was the known parity doublet structure of
 the excited baryons the conjecture actually implies a 
stronger constraint on the spectrum.

Effective chiral restoration  
implies that the physical states fall into chiral multiplets 
of nearly degenerate states. 
Let us consider in some detail the structure of those multiplets.
The irreducible representations of $SU(2)_L \times SU(2)_R$ 
  may be  labeled as $(I_L,I_R)$ where $I_L$ and $I_R$ represent the 
isospin of the left- and right handed SU(2) groups.
There is an automorphism of $SU(2)_L \times SU(2)_R$,
$Q^i_L \leftrightarrow Q^i_R$, where $Q^i_{L,R}$ are the
left and right chiral charges and $i$ refers to isospin. This
automorphism can be interpreted as the parity operation
$ L \leftrightarrow R$, under which the vector charge
$ Q^i = Q^i_L + Q^i_R$ is not affected, but the axial charge
$Q^i_{5} = Q^i_L - Q^i_R$, changes its sign.

QCD with $\theta=0$ respects parity; thus chiral multiplets must be
 large enough so as to contain states of good parity.  This
 can only happen if the representation transforms into itself
 under parity---{\it i.e.} under parity every state in the 
representation transforms into another state in the
 representation.  However, in general, the irreducible
chiral representations 
do not  trsansform into themselves under parity.  A general 
irreducible chiral representation, $(I_a,I_b)$ transforms under parity into
 $(I_b,I_a)$, {\it i.e.} it cannot be ascribed any definite parity 
except for those ones $(I,I)$ that transform into themselves.  
Thus, if chiral symmetry 
is effectively restored for a class of states the multiplets
 must either be chiral multiplets of the form $(I,I)$ or the
 multiplets  must be combined parity-chiral multiplets
 containing two irreducible chiral representations,
 {\it i.e.} $(I_a, I_b) \oplus (I_b,I_a)$.  Moreover, baryons 
in two flavor QCD cannot fall into $(I,I)$ chiral representations
 since all states in these representations are of integral 
isospin while baryons in two flavor QCD are all of half 
integral isospin.  Thus the effective chiral restoration 
for baryons states implies that they fall into parity-chiral
 representation of the form $(I_a, I_b) \oplus (I_b,I_a)$ 
with $I_a$ half integral and $I_b$ integral.

Now let us look at the parity structure of these parity-chiral
 multiplets.  Obviously, they contain parity-doublets.  For 
every state of good parity in the multiplet there is another
 state which has the same total isopin but opposite parity. 
 The reason for this is quite clear.  All of these parity-chiral
 multiplets contain two distinct irreducible chiral representations which 
transform into each other under parity.  A state of good parity
 can be constructed starting from a state of good isospin 
 in the $(I_a,I_b)$ representation which we denote $|I_{(I_a,I_b)}\rangle$.
  The states of positive and negative parity are 
$2^{-1/2} \left ( |I_{(I_a,I_b)}\rangle +  P |I_{(I_a,I_b)}\rangle \right )$ 
and 
$2^{-1/2} \left ( |I_{(I_a,I_b)}\rangle -  P |I_{(I_a,I_b)}\rangle \right )$
 respectively.  Thus, as advertised, effective chiral restoration 
for this class of states explains the parity doublets.  However,
 in general the parity doublet states of the given isospin
are not the only states in 
the representation that includes states of different isospin.  
Thus, in general, one expects that
 approximate degeneracy of states associated with effective
chiral restoration to include more states than simple parity
 doublet of the given isospin.

Let us now consider in some detail the phenomenological 
consequences of effective chiral restoration for states
 high in the baryon spectrum. As discussed above, such 
states would have to fall into parity-chiral multiplets 
with representations of the form $(I_a, I_b) \oplus (I_b,I_a)$
 with $I_a$ half integral and $I_b$ integral.  States in such
 representations can have isospins ranging from a maximum of
 $I=I_a + I_b$ to a minimum of $I=|I_b - I_a|$.  Empirically, 
there are no known baryon resonances which have an isospin greater
 than 3/2.  In the language of the constituent quark model this 
is equivalent to the statement that there are no known quantum-number
 exotic baryons.  Thus we have a constraint from the data that 
if chiral symmetry is effectively restored for very highly
excited baryons, the only possible representations for the
 observed baryons have $I_a + I_b \le 3/2$, {\it i.e.} the
 only possible representations are 
$(1/2,0) \oplus (0,1/2)$, $(1/2,1) \oplus (1,1/2)$
 and $(3/2,0) \oplus (0,3/2)$. Since chiral symmetry and 
parity do not constrain the possible spins of the states 
these multiplets can correspond to states of any fixed spin.  
The $(1/2,0) \oplus (0,1/2)$ multiplets contain only isospin 
1/2 states and hence correspond to parity doublets of nucleon 
states (of any fixed spin).  Similarly, $(3/2,0) \oplus (0,3/2)$
 multiplets contain only isospin 3/2 states and hence correspond 
to parity doublets of $\Delta$ states (of any fixed spin).  
However, $(1/2,1) \oplus (1,1/2)$ multiplets contain both 
isospin 1/2 and isospin 3/2 states and hence correspond to
 multiplets containing both nucleon and $\Delta$ states of 
both parities and any fixed spin.

Summarizing, if $(1/2,0) \oplus (0,1/2)$ and $(3/2,0) \oplus (0,3/2)$
were realised in nature, then the spectra of highly excited
nucleons and deltas would consist of parity doublets. However,
the energy of the parity doublet with  given spin in
the nucleon spectrum {\it a-priori} would not coincide with the
energy of the doublet with the same spin in the delta spectrum.
This is because these doublets would belong to different
representations of $SU(2)_L \times SU(2)_R$. On the other hand,
if $(1/2,1) \oplus (1,1/2)$ were realised, then the highly
lying states in $N$ and $\Delta$ spectrum would consists of
multiplets that contain one $N$ parity doublet and one $\Delta$
parity doublet with the same spin and are degenerate in mass.
We stress that this classification is the most general one
and does not rely on any model assumption about the
structure of baryons.

\section{Can $U(1)_V \times U(1)_A$ restoration explain parity doublets?}

Before discussing the data in detail it is useful to 
consider briefly an alternative explanation for parity
 doublets in the spectrum namely effective 
$U(1)_V \times U(1)_A$ restoration \cite{Duck}.
 At first sight this seems to be a more natural explanation 
of the parity doubling phenomena as it does not seem to 
require larger multiplets and involve doublets of the given
isospin only.  One might postulate, for example, 
that instanton effects responsible for anomalous $U(1)_A$ 
violations may become unimportant high in the baryon spectrum. 
 However such a scenario is highly implausible in our view.  
 $U(1)_V \times U(1)_A$ is
 broken in two ways---explicitly through the axial anomaly 
(and quark masses) and spontaneously.  It is clear that both
 types breaking are present. Consider, for example QCD with 
massless quarks in the large $N_c$ limit (or imaginary world
with $N_c=3$ but without axial anomaly).  In the large $N_c$
 limit all effects of the axial anomaly are absent. However  due 
to the spontaneous symmetry breaking the pion is massless while 
its $U(1)_A$ partner the isovector scalar $\delta$ is not.  
Indeed the same condensates which break $SU(2)_L \times SU(2)_R$
 (such as $\langle \overline{q} q \rangle$) also break 
$U(1)_V \times U(1)_A$ . 
 Thus even if one can argue that the effects of anomalous $U(1)_A$
 breaking shut off high in the baryon sector, unless the effects
 of spontaneous breaking of $SU(2)_L \times SU(2)_R$
also shut off one will not have parity
 doublets.  However, if the effects of the spontaneous 
breaking of $U(1)_V \times U(1)_A$
 do shut off one would expect that the effects of spontaneous
 $SU(2)_L \times SU(2)_R$ would also shut off as the spontaneous 
breaking of both types of symmetries involves the same condensates.

\section{Experimental data}

The question of relevance is whether the observed baryon
 highly lying resonances fall into these representations.  This is not
 trivial to determine for a number of reasons.  The first
 is that even if the conjecture is correct the effective
 chiral restoration is only approximate due to both quark mass 
effects and residual effects of spontaneous symmetry breaking.
  Moreover, we have no tools to estimate in an {\it a priori}
 fashion the expected size of these symmetry-breaking effects
 high in the baryon spectrum.  Thus some judgment is need to
 assert that two levels are ``nearly degenerate''.  A second
 complication stems from the fact that this high in the spectrum
 there are many levels close together and one cannot 
rule out the possibility that two states are near each other 
in energy by accident.  Moreover, the
experimental data \cite{PDG}
 is neither perfect nor complete in this region and the extraction of 
 resonance masses
 from the data introduces additional uncertainties.

Keeping all these in mind, we note however, that the
known empirical spectra of the highly lying $N$ and 
$\Delta$ baryons  suggest remarkable regularity.
Below we show all the known $N$ and $\Delta$ resonances in the
region 2 GeV and higher and include not only the well
established baryons (``****'' and ``***'' states according to the
PDG classification \cite{PDG}), but also ``**'' states that are defined
by PDG as states where ``evidence of existence is only fair''.
In some cases we will fill in the vacancies in the classification
below by the ``*'' states, that are defined  as 
``evidence of existence is poor'' and mark these states by (*).

$${\bf J=\frac{1}{2}:}
~N^+(2100)(*),N^-(2090)(*),\Delta^+(1910),\Delta^-(1900);$$

$$ {\bf J=\frac{3}{2}:} 
~N^+(1900),~N^-(2080),~\Delta^+(1920),~\Delta^-(1940)(*);$$

$${\bf J=\frac{5}{2}:}
 ~~N^+(2000),~N^-(2200),~ \Delta^+(1905),~\Delta^-(1930)~;$$

$${\bf J=\frac{7}{2}:}
~N^+(1990),~ N^-(2190),~\Delta^+(1950),~\Delta^-(2200)(*);$$

$${\bf J=\frac{9}{2}:}
~N^+(2220),~N^-(2250),~\Delta^+(2300),~\Delta^-(2400)~;$$

$${\bf J=\frac{11}{2}:}
 ~~~~~~?~~~~~,~N^-(2600),~\Delta^+(2420),~~~~~?~~~~~~~~;$$

$${\bf J=\frac{13}{2}:}
~N^+(2700), ~~~~~~?~~~~~,~~~~~~?~~~~~ ,~\Delta^-(2750)~~;$$

$${\bf J=\frac{15}{2}:}
 ~~~~~~?~~~~~~,  ~~~~~~?~~~~~,~ \Delta^+(2950), ~~~~~?~~~~~~~.$$

The data above suggest that the parity doublets in $N$
and $\Delta$ spectra are approximately degenerate; the typical splitting 
in the multiplets are $\sim 200$ MeV or less, which is within
the decay width of those states.  Of course, as noted
above,``nearly degenerate'' is not a truly well-defined idea. 
In judging how close to degenerate these states really are
one should keep in mind that the extracted resonance masses have 
uncertainties which are typically of the order of 100 MeV. 

Though one cannot  rule out the possibility that 
(i) the approximate
mass degeneracy between the $N$ and $\Delta$ doublets is
accidental ( then it would mean that baryons are organized
according to $(1/2,0) \oplus (0,1/2)$ for $N$ and 
$(3/2,0) \oplus (0,3/2)$ for $\Delta$ parity-chiral doublets)
we believe that
this fact supports an idea (ii) that the highly excited states
fall into approximately degenerate multiplets
$(1/2,1) \oplus (1,1/2)$. 

While a discovery of states that are marked by (?)
would support the idea of effective chiral symmetry restoration,
 a definitive discovery of states that are beyond the systematics of
parity doubling, would certainly be strong evidence against it.
The nucleon states listed above exhaust all states
(``****'',``***'',``**'',``*'') in this part of the spectrum
included by the PDG. However, there are some additional
candidates (not established states) in the $\Delta$ 
spectrum. In the $J=5/2$ channel
there are two other candidate states $\Delta^+(2000)(**)$ and
$\Delta^-(2350)(*)$; there is another
candidate for $J=7/2$ positive parity state - $\Delta^+(2390)(*)$
as well as  for $J=1/2$ negative parity
state $\Delta^-(2150)(*)$. Certainly a better exploration
of the highly lying baryons is needed.

\section{Discussion}

If our conjecture is correct, and assuming that
the correlator of three quarks does couple to these states
({\it a-priori} one cannot rule out the possibility that
these states are not strongly coupled to the three quark correlator,
 but
do couple to the  correlator that contains 5 quark fields, etc)
it would imply that these highly excited   baryons behave as
though they were  made out of
two left   and one right quark fields (and vice versa).

The conjecture of ``effective chiral restoration'' with the states 
in the $(1/2,1) \oplus (1,1/2)$ representation seems to be in qualitative
agreement with the spectroscopic data.  However, 
it is essential to consider how the conjecture can be tested, {\it i.e.} 
to determine what possible types of evidence can be found which would support 
the conjecture or rule it out.
The right-left structure of these states might be in principle
studied in weak processes, but as a practical matter this 
is not possible
since the life time of these states is much below the 
typical time of weak interactions. 
In the future, if one is able to describe these states directly from QCD
the conjecture  could be checked directly, specifically one  
can test whether the states in question only couple strongly 
to currents with
$(1/2,1) \oplus (1,1/2)$ quantum numbers.

Finally we wish to discuss the relation of the present work with a scheme 
recently introduced by Jido, Hatsuda and  Kunihiro (JHK) which 
in the context of the generalized $\sigma$-model organizes
{\it low-lying} baryon fields
into the representations $(1/2,1) \oplus (1,1/2)$\cite{HATSUDA}.  In fact, 
the two schemes are quite different.  The present work is based on the 
notion that 
highly-lying
 baryons physical {\it states}  behave as if 
 they approximately form $(1/2,1) \oplus (1,1/2)$ multiplets;
  our arguments are based on the quark-hadron duality.
In contrast, the JHK scheme is based on low-lying baryon {\it fields} 
falling into such multpilets. The distinction between the symmetry
 properties of fields and of states is critical.  Of course, if the
  vacuum had not spontaneously broken chiral symmetry ( or had the
   symmetry breaking effects been very weak), 
then by acting with these fields on the vacuum one would obtain 
multiplets of 
(nearly) degenrate states in the $(1/2,1) \oplus (1,1/2)$ representation. 
 However, the vacuum {\it does} break the symmetry strongly and the
physical states in JHK  scheme are {\it not} eigenstates of chirality
and do not correspond to degenerate chiral multiplets 
 as the high-lying states do in the scheme presented here.

It is worth noting in passing that the JHK scheme  also 
implicitly
assumes that these fields when acting on the (chirally broken) vacuum produce
single narrow resonance states.   This assumption implies 
a one-to-one correspondence between the fields (which form paritiy-chiral 
multiplets) and the low-lying physical states.  While one may 
entertain this assumption as a hypothesis for the way QCD dynamics plays out,
it is not obvious {\it a priori} whether such a hypothesis can be justified. 
 It could be justified only if
there were a {\it continuous} smooth "transition" from the Wigner
mode (where the whole spectrum would consist of chiral multiplets) to Nambu-Goldstone one or if the chiral symmetry breaking
effects in the vacuum represented only a small perturbation. 
Indeed, the phenomenological data does {\it not} allow to classify all
the existing low-lying baryons into chiral multiplets and
as a consequence the well established states
 $N(1700)$,$N(1710)$,$\Delta(1600)$, and 
$\Delta(1920)$ do not fall into JHK multiplets with other known 
resonances.  On the contrary, as argued in the present paper,
the chiral symmetry breaking effects do represent only a
small perturbation at large $s$ and hence one can expect
the physical spectrum there to consist only of parity-chiral multiplets.
Thus the high lying states can fall into multiplets as
    hypothesized here without the low lying baryon fields
being organized into chiral multiplets as in the JHK scheme.

\section{Acknowledgement}

TDC  acknowledges a discussion with Robert Jaffe 
(at the workshop \cite{Duck})
who pointed out that effective chiral restoration is not 
simply equivalent to parity doubling but implies a more 
complicated multiplet strucure and suggested instead the 
possibility of $U(1)_A$ restoration.  Jaffe's argument
on $U(1)_A$ restoration was reported in  \cite{Duck}.
We are greateful to D. Diakonov, L. Kisslinger and E. Shuryak for comments
on the manuscript.
TDC acknowledges the was supported of U.S.~Department 
of Energy via grant DE-FG02-93ER-40762 and LYaG acknowledges
support from the FWF Project P14806-TPH.

\end{document}